\documentclass[aps,12pt,superscriptaddress,tightenlines]{revtex4}
\usepackage{amsmath,amssymb,amsthm,graphicx,bm,color}

\newcommand{\p}{\partial}

\renewcommand{\Delta}{\varDelta} 
\renewcommand{\Gamma}{\varGamma} 
\renewcommand{\Omega}{\varOmega} 
\renewcommand{\Phi}{\varPhi} 
\renewcommand{\Psi}{\varPsi} 
\renewcommand{\Sigma}{\varSigma} 
\renewcommand{\Theta}{\varTheta} 
\renewcommand{\epsilon}{\varepsilon}

\newcommand{\R}{\mathbb{R}}
\newcommand{\Z}{\mathbb{Z}}

\begin{document}

\title{Higher order topological actions}

\author{Roman V. Buniy}

\email{roman@uoregon.edu}

\affiliation{Institute of Theoretical Science, University of Oregon,
Eugene, OR 94703}

\author{Thomas W. Kephart}

\email{tom.kephart@gmail.com}

\affiliation{Department of Physics and Astronomy, Vanderbilt
University, Nashville, TN 37235} 

\date{\today}

\begin{abstract}
  In classical mechanics, an action is defined only modulo additive
  terms which do not modify the equations of motion; in certain cases,
  these terms are topological quantities. We construct  
  an infinite sequence of higher order
  topological actions and argue that they play a role in quantum
  mechanics, and hence can be accessed experimentally.
\end{abstract}


\maketitle

\section{Introduction} 

Measurable phases in physics, from those appearing in high energy
scattering to those on which superconducting devices are based, have
played a critical role in applying our understanding of quantum
mechanics. These phases typically arise in interference phenomena, and
can be topological or geometric. Of the topological phases, only those
related to first order (Gaussian) linking have been studied in
detail. However, there is an infinite set of higher order topological
linkings, and in this paper we argue that this higher order set has a
concomitant infinite set of phases, all in principle detectable in the
laboratory. While the proper treatment of this topic is necessarily
somewhat mathematical, the results are physically predictive and
imminently testable.

An equation of motion of a dynamical system is a stationary point of
an action. As a result, different actions can lead to the same
equation of motion. In particular, addition of a quantity $S$ to the
action does not change the equation of motion if $\delta S=0$. As an
example, consider $S=\int_C A$, where $C$ is an oriented curve and $A$
is a differential $1$-form~\cite{differential-forms}. The variation is
expressed in terms of the Lie derivative, $\delta S=\int_C\pounds_{\delta
x}A$. Since $\delta x\vert_{\p C}=0$, the condition $\delta S=0$ leads
to $dA=0$.

The condition $dA=0$ makes a set of closed curves $C$ special since
small deformations of such $C$ do not change the value of $S$. In such
a case, $S$ depends only on global properties of $C$ and consequently
it is a topological quantity. Hereafter we consider only closed curves
$C$ and call $S$ a topological term.

If $A$ is exact, then $S=0$, trivially. Hence we are interested in closed
$1$-forms which are not exact and therefore physically relavent.
Let $A^*$ denote the vector space of
such forms. We will show that $A^*=\cup_{p\ge 1}A^{(p)}$, where each
space $A^{(p)}$ is constructed from spaces $A^{(q)}$, where $q<p$. The
space $A^{(1)}$ is generated by the elements of the first cohomology
group $H^1(M)$. If $M$ is simply connected, then $H^1(M)$ is trivial
and the topological term vanishes. If $M$ is non-simply connected,
then $H^1(M)$ is nontrivial and the topological term can be
nonzero. All elements of $A^{(1)}$ are local quantities and all
elements of $A^{(p)}$ for $p\ge 2$ are nonlocal quantities; the degree
of nonlocality increases with $p$.

For a given $C$, there is an associated vector space of topological
terms, $S^*=\int_C A^*$, where $S^*=\cup_{p\ge 1}S^{(p)}$ and
$S^{(p)}=\int_C A^{(p)}$. A given closed curve $C$ belongs to one of
the homotopy classes which are the elements of the fundamental group
$G=\pi_1(M)$; the value of $\int_C A$ is the same for all curves in a
class.

In quantum mechanics, the elements of $S^*$ should form abelian
representations of the group of allowed curves. We will show that this
leads to the set of subgroups $G_*=\{G_p\}_{p\ge 1}$ such that the
elements of $S^{(p)}$ form abelian representations of $G_p$. This
means that a topological term of only one order will occur in any
action \cite{refine}. We then proceed to construct $A^{(p)}$ and $G_p$
iteratively and show their relation to the homotopy classes of paths.

While there is no reason to study topological terms in classical
dynamics, such terms are important in quantum dynamics. The
Aharonov-Bohm effect is a famous example demonstrating importance of
topological terms in quantum mechanics. We review why $S^{(1)}$ is
responsible for this effect and then argue that the higher order spaces
$S^{(p)}$, $p\ge 2$, can also lead to measurable effects in
quantum-mechanical systems.

\section{Topological terms}

Without loss of generality, as an example of a $3$-dimensional
non-simply connected space
we take $M=\R^3 - T$, where $T=\cup_{1\le i\le N}T_i$ is the union of
disjoint tubes. Each tube $T_i=C_i\times D_i$ is a direct product of a
closed curve $C_i$ and a disk $D_i$. Various topological properties of
the space $M$ can be deduced from its homology and cohomology
groups~\cite{topology}. The first homology group $H_1(M)$ is a group
of closed curves modulo those which are boundaries of surfaces. The
first cohomology group $H^1(M)$ is a group of closed $1$-forms modulo
exact forms. For the present case, the basis of $H_1(M)$ is $\{\p
D_i\}_{1\le i\le N}$ and the basis of $H^1(M)$ is $\{A_i\}_{1\le i\le
N}$. By the de Rham theorem, the $1$-forms can be chosen such that the
two bases are dual to each other, $\int_{\p D_i} A_j=\delta^i_j$.
This duality condition cannot be uniquely solved for $1$-forms; a
convenient particular solution~\cite{gauge-theory} is
\begin{align}
  A_i(x)=\int_{y\in \Sigma_i}\delta(x-y)\sum_{1\le a\le 3} dx^a *dy^a.
\end{align}
Here $\Sigma_i$ is an oriented surface for which $C_i$ is the
boundary, and $*$ is the Hodge star operator. Since $A_i$ is singular
on $\Sigma_i$ and vanishes everywhere else, a closed curve $C$ which
intersects $\Sigma_i$ once in the positive direction contributes
$\delta^i_j$ to the integral $\int_C A_j$; the duality condition
follows.
We define $A^{(1)}$ as a vector space with the basis $\{A_i\}_{1\le i\le
N}$. For a given closed curve $C$, there is an associated vector space
of first order topological terms $S^{(1)}=\int_C A^{(1)}$.

To define second order topological terms, consider $F_{ij}=A_i\wedge
A_j$ for $i\not=j$. Modulo a constant factor, $F_{ij}$ is a unique
closed $2$-form which can be expressed in terms of $A_i$ and $A_j$. We
define a $1$-form $A_{ij}$ by means of an equation
$dA_{ij}=F_{ij}$. (If $C_i$ and $C_j$ are unlinked, $\Sigma_i$ and
$\Sigma_j$ can be chosen to be disjoint, in which case $dA_{ij}=0$.) A
particular solution of this equation is
\begin{align}
  A_{ij}=\tfrac{1}{2}\gamma_i A_j -\tfrac{1}{2}A_i\gamma_j.
\end{align}
where $\gamma_i=\delta_i+\int_\Gamma A_i$ and $\delta_i$ is a
constant. A path $\Gamma$ is the part of $C$ which starts at $x_0$ and
ends at $x$; the orientations of $C$ and $\Gamma$ agree. 
We define $A^{(2)}$ as a vector space with the basis $\{A_{ij}\}_{1\le
i<j\le N}$. For a given closed curve $C$, there is an associated
vector space of second order topological terms, $S^{(2)}=\int_C
A^{(2)}$.


To proceed, we look for a closed $3$-form $F_{ijk}$ for
$i\not=j\not=k$ which can be expressed in terms of the corresponding
elements of $A^{(1)}$ and $A^{(2)}$. Without loss of generality, we
have
\begin{align}
  F_{ijk}=A_{ij}\wedge A_k +A_i\wedge A_{jk}.
\end{align}
Since $dF_{ijk}=0$, we can define quantities $A_{ijk}$ by means of
equations $dA_{ijk}=F_{ijk}$. (If the first and second order linkings
for $(C_i,C_j,C_k)$ vanish, then $(\Sigma_i,\Sigma_j,\Sigma_k)$ can be
chosen to be disjoint, in which case $dA_{ijk}=0$.) Particular
solutions of these equations are
\begin{align}
  A_{ijk}=\gamma_{ij} A_k -A_i\gamma_{jk},
\end{align}
where $\gamma_{ij}=\delta_{ij}+\int_\Gamma A_{ij}$ and $\delta_{ij}$
is a constant. 
We define $A^{(3)}$ a vector space with the basis
$\{A_{ijk}\}_{i\not=j\not=k}$. For a given closed curve $C$, there is
an associated vector space of third order topological terms,
$S^{(3)}=\int_C A^{(3)}$.

It is clear how to construct higher order topological terms. The
vector spaces $\{A^{(p)}\}$ are related to what is known in algebraic
topology as the Massey products of cohomology groups~\cite{Massey};
see also \cite{Berger}.

\section{Restrictions}

In the previous section, the spaces $A^{(p)}$ were defined only on
$M$. We now extend these definitions into the interiors of the tubes
$T$. Such extensions are always possible if certain topological
restrictions are satisfied. It turns out that in order to define the
space $A^{(p)}$, all spaces $A^{(q)}$ with $q<p$ have to be
defined. If we assume that all $A^{(q)}$ with $q<p$ are defined, then
we denote $R^{(p)}$ a set of additional restrictions needed to define
the space $A^{(p)}$. We now find $R^{(p)}$ iteratively.

No restrictions are needed to define $A^{(1)}$; this means
$R^{(1)}=\varnothing$. To find $R^{(2)}$, consider extending $A_{ij}$
inside $T_i$ for $i\not=j$. This extension is possible only if
$dF_{ij}=0$ inside $T_i$, which means that $\int_{\p T_i}F_{ij}=0$.
However, since
\begin{align}
  \int_{\p T_i}A_i\wedge A_j =\int_{T_i}d(A_i\wedge A_j)
  =\int_{C_i}A_j,
\end{align}
there is an obstruction to such a procedure unless
$\int_{C_i}A_j=0$. No new restriction is needed to extend $A_{ij}$
inside $T_j$. Therefore, $A^{(2)}$ can be defined only if a set of
restrictions
\begin{align}
  R^{(2)}=\left\{\int_{C_i}A_j=0\right\}_{ i\not=j}
\end{align}
is satisfied. This means that all pairs of distinct loops $(C_i,C_j)$
should be unlinked. To find $R^{(3)}$, consider extending $A_{ijk}$
inside $T_i$, $T_j$, and $T_k$ for $i\not=j\not=k$. Reasoning as
above, we find that $A^{(3)}$ can be defined only if a set of
restrictions
\begin{align}
  R^{(3)}=\left\{\int_{C_i}A_{jk}=0\right\}_{ i\not=j\not=k}
\end{align}
is satisfied. This means that the second order linking between any
triple of distinct loops $(C_i,C_j,C_k)$ should vanish.

It is clear how to proceed to construct higher order restrictions
$R^{(p)}$, $4\le p\le N$. In order to construct all spaces
$\{A^{(p)}\}_{1\le p\le N}$, the set of curves $\{C_i\}$ has to
satisfy the restrictions $R=\cup_{1\le p\le N} R^{(p)}$. From the
property that in order for $S^{(p)}$ to be defined, all $S^{(q)}$ with
$q<p$ have to be defined, we see that $R'=\cup_{1\le p\le N'} R^{(p)}$
is always satisfied for some $N'\le N$. In this case, all linkings of
orders $2\le p\le N'$ for $N$ curves $\{C_i\}$ vanish. For $N'=N$, the
set of curves is unlinked
through $N$th order.~\cite{linking} (As a curious
observation, note that this simplest topological arrangement of loops
provides the richest structure for the topological term.)

\section{Computation}

To derive explicit expressions for the elements of $S^{(p)}$, it is
convenient to proceed as follows. First note~\cite{topology} that for
$M=\R^3-T$, a manifold with $N$ unlinked tubes removed, the
fundamental group is $G=\pi_1(M)=\Z*\cdots *\Z$, the free
product~\cite{free-groups} of $N$ copies of $\Z$. This group is of
infinite order and it is freely generated by a set of generators
$\{a_i\}_{1\le i\le N}$. (These generators are homotopically
equivalent to $\{\p D_i\}_{1\le i\le N}$.) A generator $a_i$ is
defined as a closed path in $M$, which starts at the point $x_0$,
intersects $\Sigma_i$ once in the positive direction, does not
intersect any other $\Sigma_j$, $j\not=i$, and ends at $x_0$. The
inverse path $a_i^{-1}$ is the path $a_i$ traversed in the opposite
direction. To multiply paths, we compose them in such a way such that
the end of the previous path is the beginning of the next
path. Homotopy classes of paths are labeled by finite sets of integers
$(n_{11},\ldots,n_{N1},\ldots,n_{1l},\ldots,n_{Nl})$, and
representative paths from such classes are given by
\begin{align}
  C=a_1^{n_{11}}\cdots a_N^{n_{N1}} \cdots a_1^{n_{1l}}\cdots
  a_N^{n_{Nl}}.
\end{align}
For the topological terms $S_i=\int_C A_i$, $S_{ij}=\int_C A_{ij}$,
$S_{ijk}=\int_C A_{ijk}$ we find
\begin{align}
  S_i&=\sum_{i'} n_{ii'},\\ 2S_{ij}&=\delta_i S_j -S_i\delta_j
  +\sum_{i'j'}\sigma_{i'j'}n_{ii'}n_{jj'},\\ 4S_{ijk} &=\delta_i
  S_{jk} -S_i\delta_j S_k -S_{ij}\delta_k -\delta_i S_j\delta_k
  +2\delta_{ij}S_k -2S_i\delta_{jk}
  +\sum_{i'j'k'}\sigma_{i'j'k'}n_{ii'}n_{jj'}n_{kk'},
\end{align}
where $\sigma_{ij}=1$ for $i\le j$ and $\sigma_{ij}=-1$ for $i>j$, and
$\sigma_{ijk}=1$ for $i\le j\le k$ or $k+2\le j+1\le i$, and
$\sigma_{ijk}=-1$ otherwise. Expressions for higher order topological
terms are similarly found.

Elements of $S^{(1)}$ depend only on a path; as a result, they are
additive for multiplicative paths, $S_i(CC')=S_i(C)+S_i(C')$. In other
words, elements of $S^{(1)}$ form abelian representations of the group
$G$. The situation is different for elements of $S^{(p)}$ for $p\ge
2$; they depend on both the path and the location of the point $x_0$
through constants $\{\delta_i\}$, $\{\delta_{ij}\},\ldots$. Since the
constants can be different for different loops in a product of loops,
these topological terms are not in general additive for multiplicative
paths, but below we show that there is a particular set of terms that
are additive.

\section{Topological quantum phases}

Classical dynamics is determined by the path which extremizes the
action. In quantum dynamics, all curves (paths) $C\in G$ contribute to
an amplitude through the Feynman weight factor $e^{iS}$. This allows
interference between topologically inequivalent terms and
it means that although topological terms do not affect classical
dynamics, they can affect quantum dynamics.

If the set of restrictions $R'$ is satisfied, all spaces
$\{S^{(p)}\}_{1\le p\le N'}$ can contribute to the phase of the wave
function. This obviously presents a problem when $N'\ge 2$ since
elements of $S^{(p)}$ for $p\ge 2$ do not form abelian representations
of the group $G$. We solve this problem by constructing subgroups of
$G$ for which the abelian property of the topological terms
holds. First note that $S^{(2)}$ is independent of $x_0$ only if
$S^{(1)}$ is the zero vector space. It can be shown that in this case
a closed curve $C$ is a product of commutator loops. (A commutator
loop~\cite{free-groups} is a path $[g_1,g_2]=g_1g_2g_1^{-1}g_2^{-1}$,
where $g_i\in G$.) It is easy to verify that for the product of
commutator loops an element of $S^{(2)}$ is the sum of the
corresponding terms for each component,
$S_{ij}(CC')=S_{ij}(C)+S_{ij}(C')$. This means that elements of
$S^{(2)}$ form abelian representations of the subgroup $G_2=[G,G]$
generated by commutators of elements of $G$. It is clear that it is
enough to consider a path $C=[a_i^{n_i},a_j^{n_j}]$, for which we find
$S_{ij}=n_i n_j$.

Similarly, in order for the elements of $S^{(3)}$ to satisfy the
abelian property, $S^{(1)}$ and $S^{(2)}$ must be zero vector
spaces. In this case, the path $C$ is a product of second order
commutator loops $[g_1,[g_2,g_3]]$, where $g_i\in G$, and so elements
of $S^{(3)}$ form abelian representations of the subgroup
$G_3=[G,G_2]$ generated by commutators of elements of $G$ and
$G_2$. The simplest second order commutator loop is
$C=[a_i^{n_i},[a_j^{n_j},a_k^{n_k}]]$, for which we find $S_{ijk} =n_i
n_j n_k$.
Elsewhere~\cite{long-paper}, we will provide details and show how this
procedure for higher order topological terms $S^{(p)}$ naturally
leads to groups $G_p$ which are known as the subgroups of the lower
central series~\cite{free-groups} of $G$.

According to a theorem~\cite{path-integral} for path integrals in
non-simply connected spaces, the phase of the wave function in quantum
mechanics has to form an abelian representations of the fundamental
group. By the above construction, higher order boundary terms can be
included and the phase of order $p$ is $\int_C A$, where $A\in
A^{(p)}$ and $C\in G_p$. 

Quantum mechanics imposes restrictions on what elements
of $S^{(p)}$ are allowed to contribute to the phase. This can be seen
as follows. If a charged particle is transported along a closed curve
$C$ outside a solenoid, then its action changes by $\int_C A$, where
$A$ is the gauge potential of the magnetic field in the solenoid. The
Aharonov-Bohm effect~\cite{Aharonov:1959fk} states that the wave
function acquires a phase $\phi=\xi n\Phi$, where $\xi=e(\hbar
c)^{-1}$, $n$ is the number of times the curve wraps around the
solenoid, and $\Phi$ is the flux of the magnetic field. To relate this
to the calculation for $p=1$ above, we take a path $C=a_i^{n_i}$ with
the corresponding $S_i=n_i$, and find the first order phase
$\phi_i=\xi S_i\Phi_i$. For $p=2$, we take a path
$C=[a_i^{n_i},a_j^{n_j}]$ with the corresponding $S_{ij}=n_in_j$, and
find the second order phase $\phi_{ij}=K_2\xi^2 S_{ij}\Phi_i\Phi_j$,
where $K_2$ is a constant. Proceeding similarly, we find the phase of
order $p$,
\begin{align}
  \phi_{i_1\cdots i_p} =K_p\xi^p S_{i_1\cdots i_p}
  \Phi_{i_1}\cdots\Phi_{i_p},
\end{align}
where $K_p$ is a constant. Except for $K_1=1$, constants $K_p$ are
undetermined~\cite{constants}. We are not aware of any fundamental
quantum-mechanical principle~\cite{nonlocality} forbidding the
presence of terms with $p\ge 2$ and therefore suggest this be tested
experimentally.


Let us assume for simplicity that all $N$ loops are totally
unlinked. For $N=1$, only the usual Aharonov-Bohm term $\phi_1$ can
contribute to the phase of the wave function. For $N=2$, the second
order contribution $\phi_{12}$ is present if and only if both first
order contributions $\phi_1$ and $\phi_2$ vanish; no higher order
terms are present. The simplest generalization of the Aharonov-Bohm
effect is provided by the path $C=[a_1,a_2]$. In
Ref.~\cite{BKdetection} we proposed a test of the presence of this
term in the wave function by suggesting an experimental setup to
detect the phase which we calculated to be $\phi_{12}
=K_2\xi^2\Phi_1\Phi_2$.
%
%
%
%


We have considered the case when $M$ is $3$-dimensional. The
corresponding construction for $d=2$ can be easily obtained from the
one for $d=3$. Indeed, we can smoothly deform the tubes in such a way
that a plane intersects each tube twice along a pair of disjoint
disks; we then replace each curve $C_i$ by a pair of points. Since
$3$-forms $dF_{ij}$, $dF_{ijk},\ldots$ now vanish, there are no
topological restrictions for definitions of the spaces
$\{A^{(p)}\}_{1\le p\le N}$. All other results are readily translated
from the $d=3$ case. We will study the case $d\ge 4$
elsewhere~\cite{long-paper}.

\section{Conclusions}

The action of a system is not uniquely defined since arbitrary
topological terms can be added to the action without changing the
equation of motion. Although classical dynamics is immune to such
terms, they affect the quantum dynamics. These terms can be classified
according to their topological properties. Each term contributes a
phase to the wave function, the functional form of which is easily
distinguishable from the phases due to terms of other orders. In
particular, the phase of order $p$ is proportional to the product of
$p$ fluxes. The usual Aharonov-Bohm phase corresponds to $p=1$, and
its simplest generalization is the Borromean ring phase which
corresponds to $p=2$. Examples of higher order phases $\phi_{i_1\cdots
i_p}$ due to higher order linking \cite{Kauffman}, will correspond to
general order $p$. It should not be difficult to conduct an experiment
capable of answering the question whether higher order topological
phases play a role in quantum mechanics.


\begin{acknowledgments}
 TWK thanks the Aspen Center for Physics for hospitality while this
 work was in progress. The work of RVB was supported by DOE grant
 number DE-FG06-85ER40224 and that of TWK by DOE grant number
 DE-FG05-85ER40226.
\end{acknowledgments}


\begin{thebibliography}{100}
\bibitem{differential-forms} We summarized here few relevant facts
  about differential forms. The exterior (wedge) product is
  anti-symmetric, $dx^a\wedge dx^b =-dx^b\wedge dx^a$. The exterior
  derivative of a $1$-form $A=A_a dx^a$ is a $2$-form $dA=(\p A_a/\p
  x^b)dx^b\wedge dx^a$. The exterior product of two $1$-forms is a
  $2$-form, $A\wedge B=A_a B_b dx^a\wedge dx^b$. Higher order forms
  are obtained by repeated use of operations of differentiation and
  multiplication. For a $d$-dimensional manifold $M$ and $(d-1)$-form
  $Q$, the Stokes theorem states that $\int_{\p M}Q=\int_M dQ$, where
  $\p M$ is the boundary of $M$.
\bibitem{refine} We could write a compound multisystem lagrangian
 where this is not the case, but for the discussion here we consider a
 minimal lagrangian where only one boundary term is allowed.
\bibitem{gauge-theory} In a gauge theory this corresponds to a choice
    of gauge. The reader might find it useful to view $\{A_i\}$ as
    gauge potentials of the magnetic fields confined to the tubes
    $\{T_i\}$, but the results in the text are independent of this
    analogy.
\bibitem{linking} To be more precise, consider a tube $T_i$. If it has
  the $p$th order linking with the set of tubes
  $\{T_i\}_p^{'}=\{T_1,\ldots,T_{i-1}, T_{i+1},\ldots,T_p\}$, then it
  cannot have lower order linking with any subset of these
  tubes. Furthermore, the fact that $T_i$ has the $p$th order linking
  with $\{T_i\}_p^{'}$ means that the set $\{T_i\}_p
  =\{T_1,\ldots,T_p\}$ cannot be involved in any $q$th order linking
  for $q>p$. The Whitehead link of two tubes has non-gaussian linking
  (order $>2$).
\bibitem{constants} There is an argument allowing to calculate the
constants $K_p$ for $p\ge 2$. From the Aharonov-Bohm result, if
$(2\pi)^{-1}\xi\Phi_i\in\Z$, then the phase $\phi_i$ is
unobservable. If this is also the case for the higher order phases,
then we find $K_p=(2\pi)^{-p+1}$. In any case, the values of $K_p$ for
$p\ge 2$ should be determined by an experiment.
\bibitem{nonlocality} One possible objection that can be raised
regarding the higher order terms is that all elements of $A^{(p)}$ for
$p\ge 2$ are nonlocal quantities. After addition of these terms, the
coordinate and momentum operators are still local, but the hamiltonian
operator becomes nonlocal. This nonlocality, however, has no local
consequences. (In the magnetic field analogy, the only measurable
effect is the force acting on the particle and it is absent outside
the tubes.) This is analogous to the first order term having no local
consequences despite being the local operator
itself. Elsewhere~\cite{long-paper}, we will elaborate on this matter.
\end{thebibliography}
\end{document}